\documentclass[final,3p,twocolumn]{elsarticle}

\usepackage{lineno,hyperref}
	\hypersetup{colorlinks=true,pdfborder=0 0 0}
\usepackage{subcaption}
\usepackage{amsmath}
\usepackage{amssymb}
\usepackage{gensymb}
\usepackage{cleveref}
\usepackage{siunitx}
\usepackage[dvipsnames]{xcolor}
\modulolinenumbers[5]
\usepackage{graphicx}
\usepackage{xspace}

\usepackage{fancyhdr}

\usepackage{etoolbox}
\makeatletter
\patchcmd{\ps@pprintTitle}
  {Preprint submitted}
  {Postprint of article submitted}
  {}{}
\makeatother

\journal{Metallurgical and Materials Transactions A}

\begin{document}

\begin{frontmatter}

\title{SR-CLD: spatially-resolved chord length distributions for statistical description and visualization of non-uniform microstructures}


\address[az-am]{Graduate Interdisciplinary Program in Applied Mathematics, University of Arizona, Tucson, AZ 85721, USA}
\address[az-mse]{Department of Materials Science and Engineering, University of 
Arizona, Tucson, AZ 85721, USA}

\author[az-am]{Sheila E. Whitman}
\author[az-am,az-mse]{Marat I. Latypov\corref{cor1}}
\cortext[cor1]{corresponding author}
\ead{latmarat@arizona.edu}

\begin{abstract}

This study introduces the calculation of spatially-resolved chord length distribution (SR-CLD) as an efficient approach for quantifying and visualizing non-uniform microstructures in heterogeneous materials. SR-CLD enables detailed analysis of spatial variation of microstructures in different directions that can be overlooked with traditional descriptions. We present the calculation of SR-CLD using efficient scan-line algorithm that counts pixels in constituents along pixel rows or columns of microstructure images for detailed, high-resolution SR-CLD maps. We demonstrate the application of SR-CLD in three case studies: on synthetic polycrystalline microstructures with known and intentionally created uniform and gradient spatial distributions of grain size; on non-uniform microstructures from welding simulations; and on experimental images of two-phase microstructures of additively manufactured Ti alloys with significant spatially non-uniform distributions of laths of one of the phases. \\

\noindent Note: \textcolor{red}{this is an author-generated postprint} of the article by Whitman \& Latypov \href{https://doi.org/10.1007/s11661-025-07963-6}{published} in {\it{Metallurgical and Materials Transactions A}}. \\ DOI: 10.1007/s11661-025-07963-6

\end{abstract}


\end{frontmatter}

\pagestyle{fancy}
\fancyhf{}
\fancyhead[LO]{Postprint of \href{https://doi.org/10.1007/s11661-025-07963-6}{Whitman \& Latypov, Metallurgical and Materials Transactions A (2025)}}


\section{Introduction}
\label{sec:intro}

Process--microstructure--property relationships of materials serve as the cornerstone of materials science and engineering. Efficient materials design requires not only qualitative but also {\it{quantitative}} understanding of these relationships. Quantitative process--microstructure--property relationships require rigorous description of the materials microstructure. Rigorous quantitative description of microstructures is not trivial, especially in structural metals and alloys that have a rich variety of microstructure constituents of interest at multiple length scales \cite{kalidindi2015hierarchical}. In many structural alloys, phases and grains are the constituents of special interest as they play a decisive role in a suite of engineering properties (e.g., stiffness, strength, fatigue, toughness) \cite{fullwood2010microstructure}. 

Microstructures are typically described by the statistics of size metrics of constituents such as areas, equivalent diameters, or intercepts (chords) \cite{Donegan2013}. Areas are often considered in microstructure maps where individual constituents can be clearly isolated: e.g., electron back-scattered diffraction (EBSD) maps \cite{Toth2013} or segmented optical/electron microscopy images. It is common practice to convert areas of constituents (especially grains) into {\it{equivalent}} diameters, e.g., diameters of circles of the same area as the constituent \cite{polonsky2018defects}. The equivalent diameter is often a more preferred geometric descriptor than the area even for irregularly shaped grains because it is intuitive and compatible with widely used property models, e.g., the Hall--Petch model relating the yield strength to the average grain diameter of polycrystalline metals and alloys \cite{hall1954variation,petch1956xvi}. For significantly non-equiaxed microstructures (e.g., in rolled alloys with  elongated grains), equivalent ellipses can be considered instead of circles \cite{Saylor2004,Schwartz2009}. The ellipse representation allows analyzing distributions of major and minor diameters as well as aspect ratios and inclination angles of the major axes \cite{Hovington2009}, which provides insights into not only size but also, to some extent, morphology of the constituents and their geometric orientations.   

\begin{figure*}[!h]
  \centering
  \includegraphics[width=\textwidth]{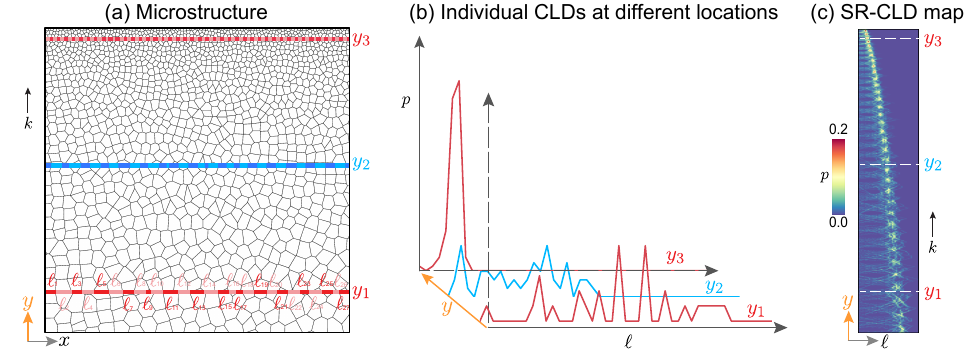}
  \caption{Illustration of the SR-CLD calculation: (a) polycrystalline microstructure with overlaid horizontal chords obtained for selected vertical locations; (b) distributions of horizontal chords (CLDs) calculated at three locations along the $y$ axis; (c) SR-CLD map as a stack of individual CLDs with the probability density, $p$, represented by color.}
  \label{fig:explanation}
\end{figure*}

The intercept, or {\it{chord}}, is another size metric used in statistical microstructure analysis for both equiaxed and non-equiaxed constituents \cite{hilliard2003stereology, underwood1969stereology,underwood1970quantitative, russ2012practical, abrams1971grain}. A chord is a line segment completely contained within a microstructure constituent (see \Cref{fig:explanation}(a)). The chord has the following advantages. First, chords can be uniquely defined and measured for constituents of arbitrarily complex shapes (and independent of their convexity or concavity \cite{underwood1969stereology}) without approximating them to a circle, ellipse, or any other idealized shape. Second, multiple important and 3D microstructure descriptors (surface area per unit volume, ASTM grain size) are directly related to the mean chord using stereological relationships \cite{underwood1969stereology}. Finally, chord lengths are relevant to transport, optical, and other properties in heterogeneous materials \cite{Torquato1993,malinka2014light,Roberts1999} and properties dictated by free paths in the microstructure, e.g., slip resistance related to the dislocation free path between grain boundaries \cite{Adams2012,Fromm2009,Sun2012}. 

In standardized practice, chords are sampled using test lines (also known as linear probes \cite{hilliard2003stereology}) or other simple test objects (e.g., circles) \cite{astm-e112}. To this end, one randomly (or systematically) overlays a test line (or a set of test lines)  with the microstructure map and then identifies intersections with the constituents of interest or their boundaries \cite{astm-e112}. Chord lengths can be then estimated from the number of intersections per test line of a known length \cite{astm-e112,hilliard2003stereology, underwood1969stereology,underwood1970quantitative, russ2012practical}. Upon sampling, the mean chord length can be obtained and reported either directly as a lineal measure of the constituents size \cite{underwood1969stereology} or, in the case of grain analysis, converted to the ASTM grain size using standardized tables or using stereological relationships the tables are based on \cite{hilliard2003stereology,underwood1969stereology,underwood1970quantitative, russ2012practical}. 

Some aspects of these standardized protocols of chord length analysis are rooted in the historically manual measurements of non-digital micrographs. The emergence of image processing, computational statistics, and visualization, as well as a shift towards digital microstructure data from contemporary instruments, present an opportunity to revisit these practices exploring not only automation but also new, more informative analyses and their visual presentations. First of all, for digital microstructures, chord lengths can be ``measured'' directly (e.g., by pixel counting \cite{hilliard2003stereology,russ2012practical,turner2016efficient} or as distances between intersections with boundaries \cite{Latypov2017}), rather than estimated from the number of intersections (critical for speed of manual or semi-manual measurements). Second, automated and direct calculation allows obtaining chord lengths from a large number of test lines overlaid with the microstructure, as opposed to sampling with only one or a few randomly placed test lines. Randomly placed test lines ensure independence of ``observations'' of chord lengths along relatively few test lines (inevitable in manual analysis) for rigorous statistical analysis. At the same time, properly random test lines (with equiprobable orientations and uniform distribution along axes) ensure the invariance of chord length distributions under rotations and translations, which is once again applicable for statistically isotropic and uniform microstructures \cite{hilliard2003stereology}.   

On the other hand, systematic sampling with a dense set of parallel lines allows detailed and quantitative analysis of constituent sizes in anisotropic (``oriented'') and non-uniform microstructures. For anisotropic microstructures, conventional intercept analysis can provide the mean chord length or ASTM grain size for a given direction in the microstructure based on intersection of constituent boundaries with an array of parallel test lines placed in that direction. The ASTM standard discusses such analysis along principal directions as well as at \ang{0}, \ang{45}, \ang{90}, and \ang{135} angles with respect to a reference direction \cite{astm-e112}. Such analysis can be further extended to all directions, which results in a polar plot referred to as the rose \cite{underwood1970quantitative,russ2012practical,saltykov1958stereometric}. To capture further details of chord length \emph{distributions} rather than their mean values or other moments, researchers recently introduced angularly-resolved chord length distributions (AR-CLDs), which can be calculated for constituent boundary maps (especially suitable for EBSD) \cite{Latypov2017} or grid-based microstructure maps (both in 2D and 3D) \cite{turner2016efficient}. For  calculation of chords in an arbitrary direction in three dimensions, Turner et al.\ \cite{turner2016efficient} used the Bresenham's method of approximating lines in raster images \cite{Bresenham1965}.

Spatially non-uniform distribution of constituents is another case of the departure from statistically homogeneous and isotropic microstructures that requires special treatment of constituent sizes. Perhaps for the reason that non-uniform (e.g., gradient) microstructures are historically less commonly studied than oriented structures, quantitative analysis and visualization of constituent sizes is underdeveloped. One existing approach is to digitally measure the sizes of individual constituents and then assign a color to each constituent according to its size (see MTEX \cite{PlottingMTEXa}). Another approach reported in literature is the calculation and visualization of moving averages of a selected size metric \cite{Lehto2016,baudoin2016numerical,russ2012practical}. Moving averages are suitable for capturing spatial size variation along a chosen direction in the microstructure, however, their use of the mean value (or other statistical moment) implicitly assumes that the chord length (or other size metric) fits a particular distribution. While much research has been published on statistical distributions suitable for describing materials microstructures, even idealized microstructures (e.g., matrix filled with second-phase spheres) may not perfectly fit the chosen distribution function \cite{gerlt2021non}. 

The selection of a suitable statistical description is even more challenging for complex microstructures that are increasingly present in materials with the emergence of new paradigms of spatially-resolved microstructure control. Examples include natural and engineering heterostructured \cite{zhu2023heterostructured,romero2025paradoxes}, architectured \cite{estrin2021architecturing}, gradient \cite{wei2014evading}, or lithomimetic \cite{beygelzimer2021earth} materials as well as materials obtained with such non-uniform synthesis or processing methods as friction stir welding \cite{heidarzadeh2021friction}, additive manufacturing \cite{hu2023functionally}, severe plastic deformation \cite{kang2016multiscale}, or epitaxial growth \cite{lo2002new}. These developments demand new methodologies of spatially-aware quantitative analysis and visualization of constituent sizes. New approaches and tools are further needed for quantitative description of digitally created microstructures, whether from physics-based simulations \cite{mitchell2023parallel}, synthetic generators \cite{Groeber2014,quey2011large}, or generative machine learning models \cite{hoffman2025grainpaint}.

Building on the prior work of moving averages and directionally resolved CLDs, we present a method of the calculation and visualization of high-resolution spatially-resolved chord length distributions (SR-CLDs). Our method is developed to capture spatial variations of the constituent size distributions along selected directions in diverse types of microstructures. \Cref{sec:srcld} details our SR-CLD methodology and then \Cref{sec:case} demonstrates its application in three case studies. 

\section{Spatially-resolved chord length distributions}
\label{sec:srcld}

\subsection{Theoretical background}
\label{sec:theory}

Our SR-CLD methodology is based on a classical definition of the CLD, also known as the chord length probability density function and denoted as $p(\ell)$ \cite{torquato2002random}. With CLD defined as a probability density function, it follows that \cite{torquato2002random}:

\begin{enumerate}
    \item $p^{(h)}(\ell)d\ell$ represents the probability of encountering a chord of length between $\ell$ and $\ell+d\ell$ in phase $h$ of the microstructure. 
    \item CLD is non-negative, i.e., $p^{(h)}(\ell)\ge0$ for all $\ell$  and $p^{(h)}(\ell)d\ell$ integrates to unity:
\end{enumerate}

\begin{equation}
\int_0^{\infty} p^{(h)}(\ell)d\ell = 1.
\label{eqn:pdf}
\end{equation}

Traditionally, thus defined CLD is estimated over the entire area or volume of a statistically uniform microstructure. Our methodology adapts this definition to account for local variations by computing CLDs at specific locations and along specified directions in microstructures of even statistically non-uniform materials. Consider a microstructure exhibiting a spatial variation in constituent sizes along an arbitrary direction $\mathbf{r}$. To capture this variation, we introduce a dense set of parallel test lines that are normal to $\mathbf{r}$. Each test line samples chords at the corresponding specific location along the $r$ axis. The ensemble of CLDs from all these locations constitutes the SR-CLD as defined in this study. Thus, SR-CLD quantifies the probability of finding a chord of a given length range within a selected phase along a specific direction and at a specific location (along the axis normal to the chosen direction).

\subsection{Computation in digital microstructure maps}
\label{sec:calc}

Following the conceptual introduction in \Cref{sec:theory}, we now present a practical method for computing SR-CLDs for digital microstructure maps. Although we illustrate the procedure for microstructure images, our methodology applies to any map on a regular grid with pixels carrying labels that distinguish the constituents of interest. We further focus our description and examples to principal axes of microstructure maps, however, the presented method can be extended to arbitrary directions either using Bresenham lines for grid-based maps \cite{turner2016efficient} or intersection-based measurements for boundary maps \cite{latypov2018application}. In what follows, we explicitly indicate for which phase the SR-CLDs are computed so that we omit the phase superscript ($h$ in $p^{(h)}(\ell)$) and instead use superscripts to indicate the spatial location.

Consider a digital microstructure map represented as a grid of pixels, where each pixel value uniquely identifies a constituent. An example is a binary image of a single-phase polycrystalline material (\Cref{fig:explanation}(a)), where grains and grain boundaries are represented by distinct pixel values (e.g., 0 and 1). When analyzing chords aligned with principal axes of the grid, spatial locations are identified by pixel indices: column indices for horizontal and row indices for vertical locations. To compute a location-specific CLD, we ``measure'' chord lengths along an individual row or column using the scan-line method \cite{turner2016efficient}, which systematically traverses the pixel grid to identify continuous sequences of pixels belonging to the constituent of interest. For example, when scanning a row, the method counts contiguous pixels with identical labels corresponding to the target constituent. Each time a different label (e.g., other phase or boundary) is encountered, the current count is recorded as a chord length (in pixels), and counting resumes from the new segment. After scanning the entire row, the pixel length are converted into physical lengths (e.g., in \SI{}{\micro\meter}) according to the grid resolution and then the physical chord lengths are binned to estimate a row-specific distribution of horizontal chords using the following expression:

\begin{equation}
p_i^{(k)}(\ell) = \cfrac{N_i^{(k)}}{\Delta \ell \displaystyle\sum_{j=1}^B N_j^{(k)}}, 
\label{eqn:cld}
\end{equation}

\noindent where $k$ denotes the index of the pixel row, the index $i$ enumerates bins from $1$ to $B$, and $N_i$ denotes the number of chords within the interval $\Delta \ell$ centered at the $i$th bin. \Cref{fig:explanation} illustrates this procedure for three scan lines (\Cref{fig:explanation}(a)) and their corresponding CLDs (\Cref{fig:explanation}(b)). If the microstructure map consists of $K$ rows, this process yields $K$ sets of chord lengths and thus $K$ location-specific CLDs. The computation of chord lengths and their distributions in the vertical direction follows the same procedure. The only difference is that pixel columns are (vertically) scanned, instead of rows. For each $k^\text{th}$ row or column in the microstructure map, the corresponding location-specific CLD satisfies the discrete normalization condition analogous to the continuous one in \Cref{eqn:pdf}:
 
\begin{equation}
\displaystyle\sum_{i=1}^B  p_{i}^{(k)} \Delta \ell = 1. 
\label{eqn:pmf}
\end{equation}

\begin{figure*}[!h]
  \centering\includegraphics[width=\textwidth]{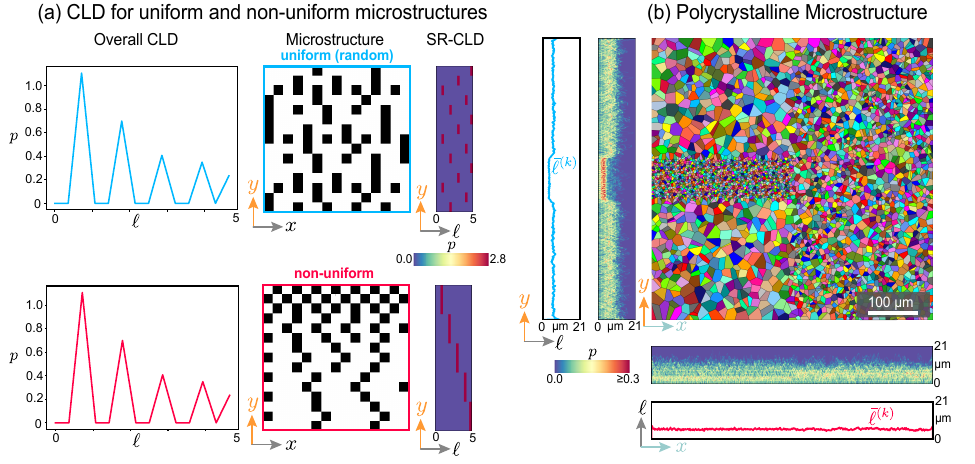}
  \caption{(a) Illustration how overall size distributions (here, overall CLDs) can be identical for two very different microstructures: (i) one with random spatial distribution of a constituent of interest (white pixels), (ii) one with non-uniform, gradient variation of the volume fraction of the constituent; and how SR-CLD can capture such differences in microstructures. (b) Demonstration of SR-CLD's ability to capture bimodal size distributions of chords throughout a synthetic polycrystal microstructure.}
  \label{fig:motivation}
\end{figure*}

Note that aggregating chord lengths from all rows or columns into a single dataset would result in the {\it{overall}} CLD for a given direction, as considered in previous studies \cite{turner2016efficient,latypov2018application}. While useful, the overall CLD overlooks spatial variation as illustrated in \Cref{fig:motivation}. 
Specifically, the overall horizontal CLDs for two synthetic microstructures are identical,  even though the microstructures are clearly different with one being spatially uniform and the other exhibiting a gradient in the spatial distribution of the constituents (\Cref{fig:motivation}(a)). It is the consideration of chord lengths along individual rows/columns followed by row-/column-wise calculation of distributions that {\it{spatially resolves}} the CLD to capture spatial variations of the microstructures. 
Mathematically, the spatial awareness of CLDs, $p^{(k)}$, is signified by the superscript $k$, which enumerates rows for horizontal CLDs or columns for vertical CLDs. Previously published AR-CLD approaches \cite{turner2016efficient,latypov2018application} focus on overall CLDs (i.e., sampled across all locations) for many directions in the microstructure but {\it{without}} spatial awareness uniquely captured by SR-CLD proposed in this study. SR-CLDs characterize the probability of finding a chord in a given range of lengths for both specific direction and specific location.
\subsection{Visualization}
\label{sec:viz}

The procedure described in \Cref{sec:calc} yields SR-CLDs, $p^{(k)}_i$, that are numerically represented by two-dimensional probability density arrays (matrices) of size $K\times B$, where $K$ corresponds to the number of pixel rows or columns in the microstructure map, and $B$ is the number of chord length bins. To intuitively represent SR-CLDs, we propose visualizing SR-CLD arrays as heat maps, where one axis represents the spatial coordinate (along which the spatial variation of CLDs is captured), the other axis represents the chord length, and the color encodes the probability density. Conceptually, such heat maps can be thought of as sequences of CLDs computed across the microstructure and stacked along the spatial axis, with the probability density axis represented by color (\Cref{fig:explanation}(c)). As illustrated in the simple example in \Cref{fig:motivation} and in subsequent case studies, such SR-CLD heat maps provide quick and intuitive visual assessment of the spatial uniformity (or lack thereof) within a microstructure. We emphasize that the SR-CLD maps presented in this study visualize raw values of the chord length probability density, $p^{(k)}_i$, without interpolation or fitting to any particular distribution function. This approach advantageously avoids the need for a priori assumptions about the functional form of the CLD, which may vary depending on the microstructure. This flexibility is evident in \Cref{fig:motivation}(b), which presents SR-CLD for a highly non-uniform polycrystalline microstructure. The left region contains patches of fine and coarse grains, whereas the right region consists of grains of consistent size. This microstructure exhibits not only variation in grain size but also in the {\it{character}} of the distribution (bimodal on the left and unimodal on the right). This example highlights the difficulty of selecting a single functional form to represent the CLD, even within a single (albeit complex) microstructure, let alone across a diverse set of microstructures.

\begin{figure*}[!h]
  \centering
  \includegraphics[width=\textwidth]{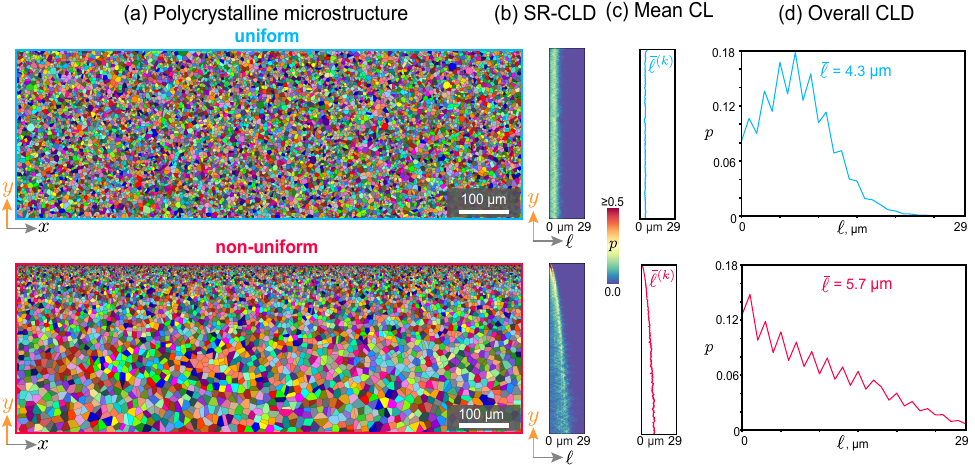}
  \caption{Description of (a) two polycrystalline microstructures with a uniform and non-uniform (gradient) distribution of grain size using (b) SR-CLD; (c) spatially resolved mean chord length (CL), (d) overall (non-spatially-resolved) CLD.}
  \label{fig:neper}
\end{figure*}

The SR-CLD values in the numerical arrays and their corresponding heat map visualizations depend on the binning of chord lengths selected in the CLD calculation (\Cref{eqn:cld}). The choice of the binning, and specifically the number of bins for calculating a distribution is not trivial; and much research has been dedicated on determining the optimal number of bins that convey a representative distribution of a given dataset \cite{sturges1926choice,scott1979optimal,freedman1981histogram,terrell1985oversmoothed,doane1976aesthetic}. Some common methods of determining the number of bins include the square root rule, Sturges' formula \cite{sturges1926choice}, Scott's normal reference rule \cite{scott1979optimal}, Freedman--Diaconis Rule \cite{freedman1981histogram}, Rice's rule \cite{terrell1985oversmoothed}, and Doane's formula \cite{doane1976aesthetic}. In this work, we explored all these methods and chose Doane's formula for SR-CLD calculations due to the large number of chords and skewness of the distribution present in most considered cases. 
 
\section{Case studies} 
\label{sec:case}

We demonstrate the application of our methodology in three case studies: (i) synthetic polycrystalline microstructures, (ii) microstructures from welding simulations, and (iii) experimental two-phase microstructures. The first case study serves as a proof of concept in which SR-CLD describes a known and intentionally created gradient in the grain size in synthetically generated polycrystals. The second case study investigates SR-CLD description of highly non-uniform microstructures simulated for a pulsated weld pass. The third case study analyzes non-uniform microstructures experimentally obtained in additively manufactured Ti alloys. Computation and visualization of SR-CLDs follows the overall procedure described in \Cref{sec:srcld}, the details of data processing specific to each case study are given below.

\subsection{Synthetic polycrystalline microstructures} 
\label{sec:neper}

To evaluate the SR-CLD approach on a microstructure with known non-uniform spatial distributions of the constituent size, we generated synthetic 2D polycrystals using the open-source software Neper \cite{quey2011large}. We generated two representative polycrystals (\Cref{fig:neper}(a)): (i) a polycrystal with uniform grain size and (ii) a polycrystal with a monotonically decreasing grain size along the vertical direction (the $y$ axis in \Cref{fig:neper}). To this end, we used different settings of grain seeding available in Neper: random for the uniform polycrystal and biased for the polycrystal with a grain size gradient. The biased seeding aimed for a \SI{45}{\percent} increase in grain size from top to bottom ends of the microstructure. Both polycrystals contained \SI{10136}{} grains and comparable mean grain size (\Cref{fig:neper}(d)).

To quantitatively compare spatial variations in these polycrystals using the SR-CLD approach, we  exported the microstructures generated in Neper as binary images. In the binary images of size $2112\times6345$ (in pixels), grain interiors were represented with white and grain boundaries with black pixels. By counting uninterrupted stretches of white pixels in each row, we digitally measured horizontal chords and computed $K=2112$ CLDs for all of 2112 pixel rows using \Cref{eqn:cld} with $B=52$ bins, as determined from Doane's formula. \Cref{fig:neper}(b) displays the resulting SR-CLD visualized as heat maps. 

\begin{figure*}[!h]
  \centering
  \includegraphics[width=\textwidth]{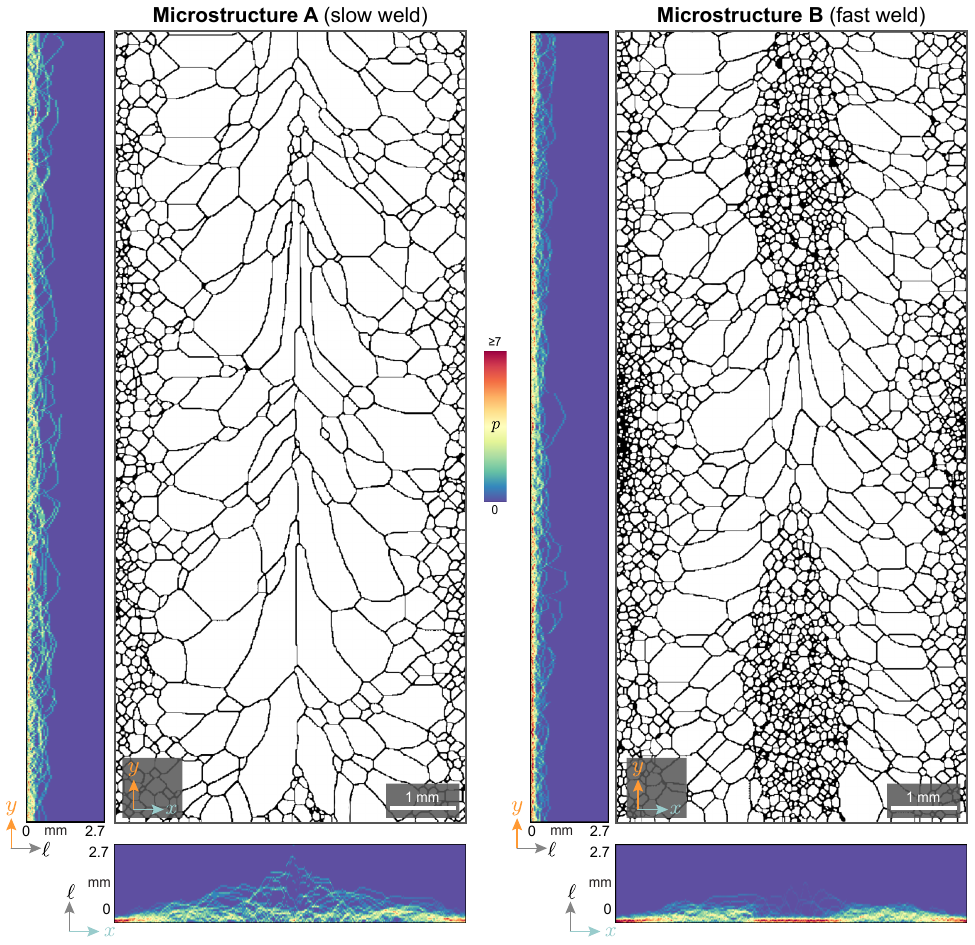}
  \caption{Two polycrystalline microstructures from simulations of pulsed welding \cite{rodgers2017monte} obtained at different welding speeds and their vertical and horizontal SR-CLDs.}
  \label{fig:spparks}
\end{figure*}

The SR-CLD clearly captures the gradient in the grain size when it is present: the SR-CLD for the gradient polycrystal shows a shift of the prevalent chord length from about 0.2 to \SI{14}{\micro\meter} (\Cref{fig:neper}(b)). This trend is confirmed with the moving average chord length shown in \Cref{fig:neper}(c). Unlike the mean chord length, however, the SR-CLD map additionally shows the variance in the chord lengths for each vertical location. The top part of the microstructure has small grains of consistent size, as seen from a high probability density in the narrow range of chord lengths (red bands in the SR-CLD map, \Cref{fig:neper}(b)). On the other hand, the probability density is lower and spread over a wider range of chord lengths for the bottom part of the gradient microstructure containing large grains with a greater variety of horizontal chord lengths. In contrast to the gradient polycrystal, the microstructure generated with random seeding is characterized by a consistent mean chord length (\Cref{fig:neper}(c)) and a SR-CLD with no trend: the chord length probability density is in a consistent range centered around \SI{4}{\micro\meter} (\Cref{fig:neper}(b)).

\subsection{Simulated welded microstructures} 

Our second case study analyzed microstructures from physics-based simulations of welding by Rodgers et al.\ \cite{rodgers2017monte}. These simulations utilize the Potts Monte-Carlo model and capture melting, solidification, and subsequent evolution of the polycrystalline microstructure in the fusion and heat-affected zones of a weld under different process conditions (e.g., weld speed).   

In this case study, we considered two simulation results obtained for pulsed welding with low ({\it{microstructure A}}) and high ({\it{microstructure B}}) speeds of the welding spot \cite{rodgers2017monte}. Welding speed affects the competition between epitaxial growth and nucleation and thus results in distinct microstructures (large irregular grains vs.\ fine equiaxed grains). To describe these differences, we pre-processed RGB images (as presented by Rodgers et al.) into the binary format consistent with the scan-line algorithm of measuring chord lengths. The $2976\times1316$ images that we adopted contained distinctly colored grains enclosed by dark grain boundaries. For segmentation, we first isolated the dark grain boundaries by thresholding pixel values, then filled small missegmented regions in the grains, and manually corrected the remaining mislabeled pixels to produce binary images with white grain interiors and black grain boundaries (\Cref{fig:spparks}). For horizontal SR-CLDs capturing horizontal variation of chord lengths, we measured vertical chords and their \SI{1316}{} location-specific distributions with 44 bins (from Doane's formula) for each image. Similarly, vertical SR-CLDs consisted of \SI{2976}{} distributions of horizontal chords with 46 bins.

The results demonstrate the utility of SR-CLD maps for describing spatial variations of grain size in highly non-uniform microstructures that develop in welding (and, by analogy, in additive manufacturing). The vertical SR-CLDs (on the left to microstructures in \Cref{fig:spparks}) show non-zero probability densities in a range of chord lengths indicating non-unimodal distributions. These non-unimodal distributions arise from combinations of large and fine grains seen in both microstructures. The spatial variations of large and fine grains are even more evident in horizontal SR-CLDs (below the microstructures in \Cref{fig:spparks}). In microstructure A, the fusion zone with large irregular grains leads to non-zero probability densities scattered in a broad range of vertical chord lengths, whereas fine grains on the sides result in high probability density in a small range of short chords. For microstructure B, there is an additional spatial range (along the $x$ axis) of high probability density of short chords in the fusion zone, which corresponds to patches of fine grains due to nucleation favored over epitaxial growth at high welding speeds \cite{rodgers2017monte}.

\subsection{Two-phase microstructures of titanium alloys} 
\label{sec:ti}

The third case study demonstrates the application of SR-CLD on real microstructures to describe spatial variation of phase sizes. Specifically, we quantify the spatial variation of the $\alpha$ phase in two-phase microstructures of two dissimilar titanium alloys additively manufactured and experimentally characterized by Kennedy et al.\ \cite{kennedy2021microstructure}. The authors co-deposited Ti5553 and Ti64 alloys using a wire-arc additive manufacturing process, which results in spatial variations of the composition, microstructure, and thus properties (e.g., strength and damage tolerance). The spatial microstructure variation is primarily manifested in a variation of the lath size of the $\alpha$ phase. To demonstrate our approach for these materials, we processed the raw experimental images of the phases into the binary format consistent with the scan-line method, aligned the individual images into large composite images and computed SR-CLDs for the entire composite images. 

\begin{figure*}[!h]
  \centering
  \includegraphics[width=\textwidth]{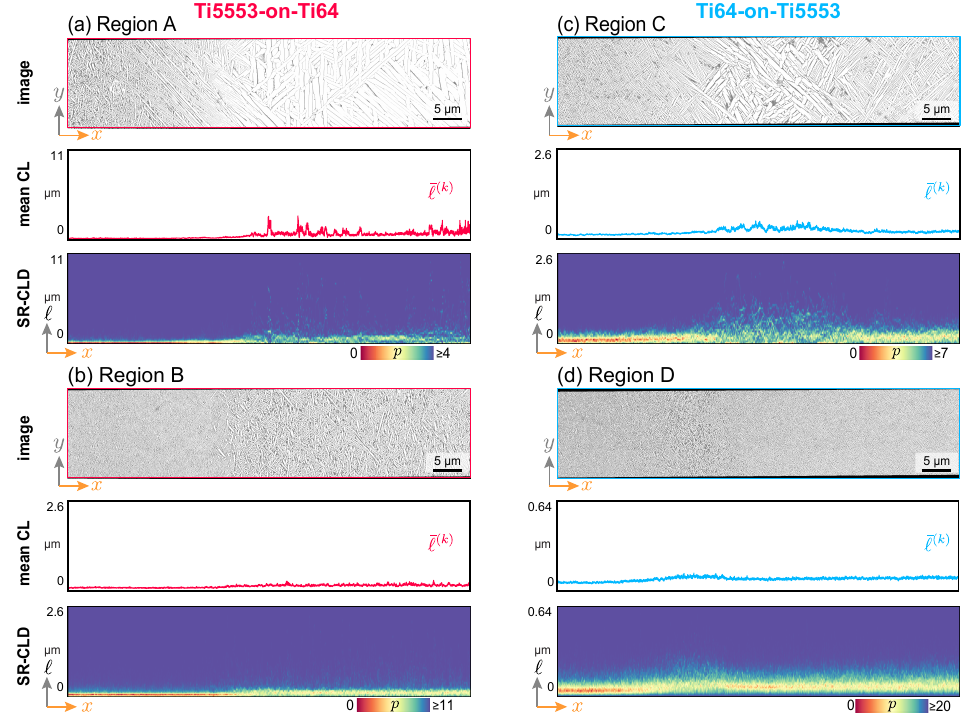}
  \caption{Description of non-uniform two-phase microstructures of additively manufactured dissimilar titanium alloys by SR-CLD and spatially-resolved mean chord lengths: (a) region of the Ti5553-on-Ti64 sample with a gradient in the size of $\alpha$ laths (white); (b) a region of the Ti5553-on-Ti64 sample with a relatively uniform microstructure; (c) a region of the Ti64-on-Ti5553 sample with a band of coarse $\alpha$ laths; (d) a region of the Ti64-on-Ti5553 sample with a relatively uniform microstructure. Note that $\alpha$ phase is shown in white here, which is inverse to images obtained by Kennedy et al.\ \cite{kennedy2021microstructure} (inverted for better presentation of results).}
  \label{fig:titanium}
\end{figure*}

Raw data published by Kennedy et al.\ \cite{kennedy2021microstructure} contain over 900 high-resolution scanning electron microscope images for two additively manufactured materials: Ti5553-on-Ti64 and Ti64-on-Ti5553. To segment the raw images, we first applied a Gaussian filter (with radius of \SI{5}{px}) and then applied a threshold selected using Yen's method \cite{366472} to obtain binary images whose white pixels represent the $\alpha$ phase (laths) of interest. The binary images were passed through erosion and dilation filters to clean up the remaining segmentation noise \cite{vincent1993grayscale, priya2017effective}. Hundreds of the individual segmented images were then aligned and merged into large composite images: one for Ti5553-on-Ti64 and one for Ti64-on-Ti5553 samples (except a couple of stained regions at the top of the characterized area). 

We computed SR-CLDs for both composite images despite their very large size ($288688\times3897$ each) following merging. Horizontal chord lengths were measured for the $\alpha$ phase by counting the corresponding white pixels at each pixel row. From the measured chord lengths, we estimated \SI{288688}{} CLDs using \Cref{eqn:cld} with $B=55$ bins, also for all \SI{288688}{} pixel rows. While we obtained the SR-CLD maps for the entire composite images, they are too large so that we present the results for  subregions of approximately \SI{70}{\micro\meter} along the $x$ axis -- the direction of the spatial microstructure variation (\Cref{fig:titanium}). \Cref{fig:titanium} shows two subregions of each processed microstructure and their corresponding SR-CLD maps and moving average chord length curves for each material. 

These subregions were selected to represent a variety of microstructure transitions that were present in the materials. The first two subregions include monotonic transitions from fine to coarse $\alpha$ laths (\Cref{fig:titanium}(a,b)), while the other two subregions feature zones of coarse  $\alpha$ laths surrounded by fine $\alpha$-lath microstructures (\Cref{fig:titanium}(c,d)). Some of these microstructure transitions are captured by the moving average chord length: e.g., coarse zone in \Cref{fig:titanium}(c,d). At the same time, the mean chord length curve has spurious peaks for the subregion shown in \Cref{fig:titanium}(a), which can mislead to a conclusion of the presence of much larger $\alpha$ laths compared to the rest of the microstructure. SR-CLD maps present a richer description of these non-uniform microstructures. Serving as visual representations of {\it{distributions}} rather than only mean values, SR-CLD maps show not only most probable chord lengths for each zone but also their consistency and variance. The high probability of short chord lengths in the zones of fine $\alpha$ laths highlights the consistency of chords in a narrow range of lengths (\Cref{fig:titanium}(a,b,d)). At the same time, the probability is distributed over a wide range of lengths for the zones with coarse laths (\Cref{fig:titanium}(a,c,d)). Partially, the lack of clear SR-CLD peaks in zones with coarse $\alpha$ laths is associated with the fewer chords present in those zones. This is because the image of (approximately) constant width captures many fine laths and relatively few coarse laths. The SR-CLD maps conveniently indicate the insufficient number of chords for conclusive statistics with low probability spread over the entire range of chord lengths analyzed for these microstructure regions. 

\section{Discussion}
\label{sec:disco}

In this work, we introduced a new approach to analyzing and quantifying non-uniform microstructures. With three case studies, we demonstrated that SR-CLDs provide insights into spatial microstructure variations inaccessible with traditional methods of microstructure descriptions. For example, two clearly different polycrystalline microstructures studied in \Cref{sec:neper} had similar overall grain size distributions and mean grain sizes (\Cref{fig:neper}(a,d)). SR-CLD captured the differences in those microstructure both numerically and visually via SR-CLD maps (\Cref{fig:neper}(b)). Capturing spatial differences is important because microstructure description is a foundation for establishing process--microstructure--property relationships. We can expect that a microstructure with a significant gradient is a result of a particular process and will have properties different from those of a uniform microstructure. Yet, microstructure descriptions that focus on overall size distributions or mean values without spatial sensitivity will fail to reflect such differences. The SR-CLD approach that resolves spatial microstructure variations can therefore serve as a statistically rigorous microstructure description for quantitative process--microstructure--property relationships in materials with significantly non-uniform microstructures. For example, SR-CLDs or their reduced-order representations (e.g., from principal component analysis \cite{latypov2018application}) could be used as microstructure representation (``features'') for machine learning, as previously shown with traditional (non-spatially-resolved) CLDs \cite{Popova2017,ackermann2023machine,fan2021quantitative}. The second and third case studies demonstrate that SR-CLDs can provide insights into spatial variations of microstructure developing in welding and additive manufacturing, which is timely given active research in the field of metal 3D printing. We also note that SR-CLD is a complementary tool to the AR-CLD methodology \cite{turner2016efficient,latypov2018application}. AR-CLD is most useful for ``oriented'' microstructures, e.g., microstructures with elongated grains or particles uniformly distributed throughout the material, whereas SR-CLD is most useful for ``gradient'' or other spatially non-uniform microstructures, such as those considered in the case studies above.

While our case studies demonstrated SR-CLDs for describing grains and phases, the presented approach is flexible for analyzing any other microstructure constituents in a wide variety of materials and data modalities. Since our chord length calculation is based on simple pixel counting, any microstructure map that contains constituent labels at pixels could be used for SR-CLD calculations. This includes electron back-scattered diffraction (EBSD) maps that contain phase IDs (for multiphase microstructures) and grain IDs at each EBSD ``pixel'', which can be used for digital chord measurements for the SR-CLD approach. Segmented microstructure images are another wide class of microstructure data that can benefit from the presented SR-CLD approach. An advantage of leveraging SR-CLDs for segmented images is that, CLDs are tolerant to segmentation errors inevitable in experimentally obtained microstructures \cite{whitman2023automated}. Since raw values of chord lengths and binned probability densities are obtained as part of the SR-CLD computation, one could easily, if necessary, fit these values to a specific probability density function (e.g., log-normal \cite{gerlt2021non}). Furthermore, since chord is fundamentally a 3D size metric sampled from 2D sections, one could further obtain standardized yet location-specific grain size for polycrystallined materials using stereological relationships \cite{underwood1970quantitative} or conversion tables \cite{astm-e112}.

Based on simple pixel counting, the presented approach is computationally efficient with minimum CPU and memory requirements. Calculation of SR-CLD for a typical high-resolution $12000\times4000$ image (shown in \Cref{fig:titanium}) takes only about \SI{23}{\second}, and \SI{6.3}{\minute} for a very large ($288688\times3897$) composite image  on an average consumer-grade laptop (MacBook Air M1 with \SI{16}{\giga\byte} RAM). Since the calculation of individual location-specific CLD for a pixel row/column is independent, the SR-CLD calculation can be easily parallelized if needed, e.g., for extremely large microstructure maps. To facilitate adoption of the approach by the community, we made a Python code for SR-CLD calculations available on GitHub (link below).

\section{Conclusion}
\label{sec:summary}

In this paper, we presented calculation of spatially-resolved chord length distribution (SR-CLD) for statistical description of non-uniform microstructures. With three case studies, we demonstrated the application of the SR-CLD approach to different microstructures with grains and phases as microstructure constituents of interest. The results show that SR-CLDs capture spatial variations of constituent sizes that would be inaccessible with traditional microstructure descriptions focusing on overall size distributions or their moments. SR-CLD captures microstructure uniformity (or lack thereof) both visually (via proposed SR-CLD maps) and numerically and are therefore suitable both for intuitive visual assessment of the microstructure and for quantitative relationships between microstructure, properties, and processing.  With a simple pixel counting algorithm as a basis, SR-CLD calculations require very modest computational resources and can be therefore calculated for description or alignment of very large images on a typical laptop.   

\section*{Acknowledgements}

SEW acknowledges the support by the National Science Foundation (NSF) Graduate Research Fellowship Program under Grant No.\ DGE-2137419. The views and conclusions contained herein are those of the authors and should not be interpreted as necessarily representing the official policies or endorsements, either expressed or implied, of the NSF.

\section*{Data availability}

The codes for SR-CLD calculations and SR-CLD-based alignment are available at \url{https://github.com/materials-informatics-az/SR-CLD}.

\section*{Conflict of interest}
On behalf of all authors, the corresponding author states that there is no conflict of interest.

\bibliography{refs.bib}{}

\end{document}